\documentclass[%
 reprint,
bibnotes,
 amsmath,amssymb,
 aps,
pra,
]{revtex4-1}
\usepackage[utf8x]{inputenc}
\usepackage[english]{babel}
\usepackage{graphicx}% Include figure files
\usepackage{dcolumn}% Align table columns on decimal point
\usepackage{bm}% bold math
\usepackage{hyperref}% add hypertext capabilities
\usepackage[mathlines]{lineno}% Enable numbering of text and display math
\usepackage{xcolor}
\usepackage{ulem} % {\sout{}}
\usepackage{natbib}

\usepackage{braket}

\usepackage{graphicx} % Required for including images
\graphicspath{{figures/}} % Location of the graphics files

\usepackage{amsfonts, amsmath, amsthm, amssymb}
\usepackage{hyperref}
\hypersetup{
    colorlinks,
    citecolor=purple,
    filecolor=purple,
    linkcolor=purple,
    urlcolor=purple
}
\begin{document}

\preprint{APS/123-QED}

\title{Statistical Entropy of Open Quantum Systems}% Force line breaks with \\

\author{L.M.M. Durão}
\email{durao@ifi.unicamp.br}
 
\author{A.O. Caldeira}%
% \email{durao@ifi.unicamp.br}
\affiliation{%
 Institute of Physics Gleb Wataghin, University of Campinas. 
}%

\date{\today}% It is always \today, today,

\begin{abstract}
 Dissipative {quantum} systems are frequently {described within} the framework of the {so-called} ``system-plus-reservoir'' approach. In this {work} we assign {their} description {to} the Maximum Entropy Formalism and compare {the resulting thermodynamic} properties with {those of} the well {-} established approaches. Due to the {non-negligible} coupling {to} the heat reservoir, these systems are non-extensive by nature, {and the former} task may require the use of non-extensive parameter dependent informational entropies. In doing so{,} we address the problem {of choosing appropriate forms of those entropies in order to describe} a consistent thermodynamics {for dissipative quantum systems}. {Nevertheless, even having chosen the most successful and popular forms of those entropies, we have proven our model to be a counterexample where this sort of approach leads us to wrong results}.


%\end{description}
\end{abstract}

\pacs{Valid PACS appear here}% PACS, the Physics and Astronomy
\maketitle

\section{\label{sec:level} Introduction}
 In recent years, quantum dissipation arose as a key concept in the design, development, and control of devices where quantum effects play a fundamental role. This new area is referred to as \textit{quantum technologies}. Quantum dissipation is also regarded as a main resource to explain the foundations of statistical mechanics\cite{popescu2006entanglement}, the rise of the  thermodynamic behaviour in quantum systems\cite{gemmer2009quantum}, and is a powerful tool to explore exotic states of matter in many body systems\cite{Budich}\cite{Diehl}. 
 
 The essence of the dissipative effects in the dynamical or equilibrium properties of quantum mechanical systems is satisfactorily captured by
 the system-plus-reservoir approach\cite{feynman1963theory}\cite{Caldeira2014}. Although these models do not answer all the existing questions related to these systems, they  provide us with a systematic technique to calculate either the dynamic or the equilibrium density operator of a system not necessarily very weakly coupled to its environment. We refer to this situation as the strongly coupled regime although this is not a very precise terminology \footnote{Actually we are interested in systems which are strongly damped even though this does not necessarily imply that the systems is strongly coupled to its environment}.

 Statistical mechanics frequently uses the idea of a small system interacting with a much larger one at a given temperature. In some cases we do not really need to construct a proper heat bath, but rather identify parts of the entire system under study (the universe) as the system (of interest) and {the remaining part as }its reservoir. For example, in a chemical reaction, we identify the molecules as a system and the solution as a reservoir. The same works for a particle in a viscous fluid (a Brownian particle)  or an atom in an optical cavity.
 
 The main issue of this approach is: we have to model not only the reservoir but also its coupling to the system. Both the form and  strength of the coupling are fundamental to identify the kind of phenomena that could appear in a dissipative framework. We are mostly interested in the strong coupling regime.

{In this regime, the traditional thermodynamic analysis seems to fail since in writing Clausius inequality, the role of the system-bath interaction in the entropy variation or exchanged heat is not clear as in the weak coupling regime. Similar problems appear when one tries to put foward the maximization of Gibbs entropy, leading us to consider a departure from our program to obtain the laws of thermodynamics microscopically or a review of its assumptions}. However, those laws are not formulated in either  {regime}, but resides in phenomenological observations and in principle should be always valid. This means that the main problem is not thermodynamics itself, but how to calculate the reduced state of the system in contact with the reservoir and correctly establish a connection with thermodynamics.
Many attempts to solve these problems have been put forward in the recent past following the algorithm proposed in \cite{CALDEIRA1983A}\cite{Caldeira1983}.
 
 Now we can ask ourselves what happens if one insists in obtaining a density operator
for the system of interest directly from a maximum entropy principle\cite{jaynes1957information}. Starting from   {an entropy functional} and {imposing some constraints to it we can} apply the Lagrange's multipliers procedure in order to obtain the density operator for a dissipative system. Dissipative systems are {generally} nonextensive (except in the extremely weakly damped case {as we will shortly see}), so such a task may require the use of generalised entropies.

{As we have recently witnessed, there has been a great effort from part of the statistical mechanics community to explain certain results which could not be explained by the application of the maximum entropy principle to the Shannon or von Neumann entropies of a given system. Instead, it is successfully done if the same principle is applied to the so-called generalised entropies which always contain an adjustable parameter that can reproduce the former well-known entropies for a specific value{\cite{Lutz}\cite{Atman}\cite{brito2015role}}. The success in some cases has led many researchers to claim that particular forms of these entropies could be regarded as a generalisation of the above-mentioned entropies{\cite{BagchiTsallis}\cite{tirnakli2016standard}\cite{moyanotsallisgellmann}\cite{TSALLIS1999}}. In this way, any deviation from the standard Boltzmann-Gibbs result could be attributed to the free parameter present in the nonextensive entropy form. In a dissipative system, for example, one would expect this parameter to be a function of the damping constant, the only new parameter introduced in our system.}

In this paper, we report our attempt to describe dissipative systems by direct maximisation of two specific forms of  generalised entropies, namely Rényi and Tsallis entropies, and use the connection with thermodynamics to calculate the properties of the system of interest and compare {them} with those results obtained from the bath of decoupled harmonic oscillators. {Actually, this acts as a test to confirm whether these entropy functions can be acceptable generalisations of the standard Shannon or von Neumann entropies. If this fails we can assure that neither of these two forms can account for the correct thermodynamic behaviour of the system. Therefore, the claim of their generality would not be correct.

 {In the example we present below we were able to show that this is really the case. The predictions of the non-extensive entropies approach is at variance with  the standard thermodynamics of our system which, on the other hand, can be achieved by more conventional and reliable methods.}

The paper is organised as {follows}: in \autoref{sec:level1} we {briefly} review {some} features of quantum Brownian motion {as developed from the coupling of a particle to } the harmonic oscillator bath and calculate {its} thermodynamic properties. In \autoref{sec:level1b} we review {the} generalised properties {obtained from a minimal model proposed for dissipative systems} based on the outcome of {the} maximum entropy procedure. In \autoref{sec:level1c} we address our conclusions and perspectives.
\section{\label{sec:level1} The Quantum Brownian Motion}

Brownian motion is the most common and useful example of a dissipative system in classical physics and also serves as a paradigm  of most models for dissipative quantum systems. Its classical dynamics is described by the well-known Langevin equation for its position or , equivalently,  the Fokker-Planck equation for its density distribution in phase space. There  are many examples of classical dissipative systems whose dynamics can be mapped onto a Langevin equation, e.g,  RLC circuits and superconducting quantum interference devices (SQUIDs).
However, given that we cannot obtain a Langevin equation from a time independent Hamiltonian or Lagrangian, we cannot use the conventional canonical quantisation procedure to study the quantum regime of the system. 
An alternative route is choose a minimal model, coupling the system of interest to a second system (the reservoir), in such a way that the classical Brownian motion is reproduced for the former in the appropriate limits . This is the ``system-plus-reservoir'' program for dissipative quantum systems.

\subsection{\label{sec:level2a}The Bath of Harmonic Oscillators}
The minimal model we have just mentioned above is usually chosen as a particle (of mass $m$ and subject to an external portential $V$) coupled to a  set of non-interacting harmonic oscillators whose Lagrangian is:

\begin{equation}\label{lagrangian}
  L= L_{{S} } + L_{{I} } + L_{{R} } + L_{{CT} },
  \end{equation}
  where, {the subscripts $S$, $I$, $R$ and $CT$ stand for system, interaction, reservoir and counter-term , respectively, and these Lagrangians read} 
  
  \begin{equation}
  L_{{S} }= \frac{1}{2}m\dot{q}^2 {-} V(q),
  \end{equation}
  \begin{equation}
  L_{{I} }= \sum_{k}C_{k}q_{k}q,
  \end{equation}
  \begin{equation}
  L_{{R} } = \sum_{k}\frac{1}{2}{m_{k}}\dot{q}_k^2 - \sum_{k}\frac{1}{2}m_k\omega_k^2q_k^2,
  \end{equation}
  and
  \begin{equation}
  L_{{CT} } = {-}\sum_{k}\frac{1}{2}\frac{C_k^2}{m_k\omega_k^2}q^2.
  \end{equation}
  {Writing the Euler-Lagrange equations corresponding to the Lagrangian above one has

  \begin{eqnarray}
  m \ddot{q} = - V^{\prime}(q) + \sum_{k} C_{k} \, q_{k} - \sum_{k}
  \frac{C_{k}^{2}}{m_{k} \, \omega_{k}^{2}} \, q , \label{eqq}\\
  \end{eqnarray}
  and
  \begin{eqnarray}
  m_{k} \, \ddot{q}_{k} = - m_{k} \, \omega_{k}^{2} \, q_{k} + C_{k}
  \, q . \label{eqqk}
  \end{eqnarray}}

{Now, taking the Laplace transform of (\ref{eqqk}) one gets
\begin{eqnarray}
\tilde{q}_{k}(s) = \frac{\dot{q}_{k}(0)}{s^{2} + \omega_{k}^{2}} +
\frac{s \, q_{k}(0)}{s^{2} + \omega_{k}^{2}} + \frac{C_{k} \,
\tilde{q}(s)}{ m_{k} \, \left( s^{2} + \omega_{k}^{2} \right)} ,
\label{lapqk}
\end{eqnarray}
which after the inverse transformation can be taken to (\ref{eqq}), yielding
\begin{eqnarray}
&& m \ddot{q} + V^{\prime}(q) +
\sum_{k} \frac{C_{k}^{2}}{m_{k} \, \omega_{k}^{2}} \,
\frac{1}{2 \, \pi \, i}
\int\limits_{\varepsilon - i \, \infty}^{
\varepsilon + i \, \infty}
\frac{s^{2} \, \tilde{q}(s)}{s^{2} + \omega_{k}^{2}} \, e^{s \, t} \,
ds \nonumber \\
&=& 
\sum_{k} C_{k}\left\{ \frac{\dot{q}_{k}(0)}{
\omega_{k}} \sin \omega_{k}t \,+ \, q_{k}(0)\cos\omega_{k}t
\right\} \label{eqqlap2}
\end{eqnarray}}
{In order to replace $\sum\limits_{k} \longrightarrow \int d\omega$ on the l.h.s. of the resulting equation of motion for $q(t)$, we
introduce the spectral function $J(\omega)$ as} 
\begin{eqnarray}
J(\omega) =
\frac{\pi}{2} \sum_{k} \frac{C_{k}^{2}}{m_{k} \, \omega_{k}} \,
\delta \left( \omega - \omega_{k} \right) ,
\label{Jw}
\end{eqnarray}
{which with the choice}
\begin{eqnarray}
J(\omega)=\left\{\begin{array}{ccc}
  \eta \, \omega & \textrm{if} & \omega<\Omega \\
  0 & \textrm{if} & \omega>\Omega, \\
\end{array} \right.
\label{ohmicJ}
\end{eqnarray}
 {where $\Omega$ stands for a high-frequency cutoff, reproduces, in the classical limit {(see \cite{Caldeira2014} or \citep{Caldeira2010} for details)}, the well-known Langevin equation}
 \begin{eqnarray}
 m \, \ddot{q} + \eta\, \dot{q} + V^{\prime}(q) = f(t) .\label{langevin2}
 \end{eqnarray}
 {Here, \,$\eta$ is the damping constant and the {instantaneous term depending on the velocity of the particle results from the integral term on the l.h.s. of (\ref{eqqlap2}) when $\Omega\rightarrow\infty$ and the fluctuating force $f(t)$ is related to the r.h.s. of (\ref{eqqlap2}) and obeys}
 \begin{eqnarray}
 &&\left\langle f(t) \right\rangle = 0 \qquad\textrm{and} \nonumber
 \\[12pt]
 &&\left\langle f(t) \, f(t^{\prime}) \right\rangle = 2 \, \eta \,
 k_{B} \, T \, \delta \left( t - t^{\prime} \right). \label{white}
 \end{eqnarray}
{The statistical properties of the fluctuating force results from the equipartition theorem as applied to the bath oscillators which allows us to evaluate the variances of their initial positions and velocities as a function of temperature.}

  Now that we have agreed on the model to be employed to treat the system, we need to decide on the formulation we will adopt to explore such a model in its quantum mechanical regime. Since the knowledge of the density operator allows us to deal with both the dynamical and equilibrium properties of our system, it is advisable we formulate our problem aiming at the full description of this very general representation of the quantum state of a system. The Feynman path integral formulation of quantum mechanics, in particular, has long proven to be the most successful method for dealing with systems strongly coupled to their environments. Below we review what will be needed for our purposes.
  
\subsection{\label{sec:level2b}Path Integral Solution}
The Feynman path integral coordinate representation of the equilibrium density operator of the whole universe, is {\cite{CALDEIRA1983A}}:

\begin{eqnarray*}
\rho(x,{\bf R};y,{\bf Q},\beta) &=& \frac{1}{\mathcal Z} \int\limits_{y,\bf Q}^{{x,\bf R}}\mathcal{D}q(\tau)\mathcal{D}{\bf R(\tau)}\times\\
&\times& \exp\left\{-\frac{1}{\hbar}S_E[q(\tau){\bf R(\tau)}]\right\},
\end{eqnarray*}
where we have used the Euclidean (imaginary time) version  of the classical action {corresponding to (\ref{lagrangian}), $\beta\equiv 1/k_{B}T$ and $\mathcal Z$ is the partition function of the whole universe}. { $x$ and $y$ ( ${\bf Q}$ and ${\bf R}$ ) stand for the coordinate representation of the particle (reservoir).} As all such quantities will be Euclidean ones, we can drop the subscript ``$E$''. 

The reduced density operator of the system is obtained performing a partial trace on the total density operator:
 \begin{equation}
  \tilde{\rho}(x,y,\beta) \equiv \int d{\bf R} \,\rho\left(x,{\bf R};y,{\bf R},\beta\right) .
 \end{equation}
 
The Feynman method\cite{feynman1998statistical} tells us that after we trace out the bath coordinates, its effects will be encoded in a kind of imaginary time ``influence functional'' ${\mathcal F}[q(\tau)]$. { Defining new variables $q'\equiv (x+y)/2$ and $q''\equiv x-y$ as, respectively, the centre of mass and relative coordinates of this thermal state \cite{Caldeira2014}, we find}

  \begin{equation*}
  \rho_{\beta}(q'',q') = \frac{1}{\mathcal Z}\int\limits_{q(0)=q'}^{q(\hbar \beta)=q''}{\mathcal D}q(\tau)\exp\left(-S_S[q(\tau)]/\hbar\right){\mathcal F}[q(\tau)],
  \end{equation*}
 where 
 \begin{equation*}
  {\mathcal F}[q(\tau)]=\oint{\mathcal D}{{\bf R}}(\tau)\exp\left\{ (-S_R[{{\bf R}(\tau)}]-S_I[q(\tau),{{\bf R}(\tau)}])/\hbar\right\}.
  \end{equation*}
The closed  {functional} integral means that it must be evaluated for paths such that ${\bf R}(0)={\bf R}(\hbar \beta)= {\bf R}$.

  For the harmonic oscillator bath, all functional integrals will be Gaussian, so we can expect that the result is a Gaussian reduced density operator with variances related to $\langle q^{2}\rangle$ and $\langle p^{2}\rangle$:
  \begin{equation}\label{redensityop}
  \rho_{\beta}(q'',q')=\frac{1}{\sqrt{2\pi\langle q^{2}\rangle}}\exp\left\{-
  \frac{(q')^{2}}{8\langle q^{2}\rangle} - \frac{\langle p^{2}\rangle}{2\hbar^2}(q'')^2\right\}.
  \end{equation}\\
{Although these variances can be directly computed from the functional integration {(see \cite{CALDEIRA1983A} for details)}, they fortunately coincide with those obtained by the direct application of the fluctuation-dissipation theorem to $\langle q^2\rangle$ and $\langle p^2 \rangle$ with the response function involved therein related to the Langevin equation (\ref{langevin2}) (see equations (\ref{varq},\ref{varp} and \ref{FDT}) below).}
\subsection{\label{sec:level2c}Thermodynamic Properties}
There is a controversy over the correct method to calculate the thermodynamic properties of a system strongly coupled to a heat reservoir. This problem has been addressed in \cite{Campisi2010}\cite{Campisi2009}\cite{Ingold2012}\cite{hanggiingoldtalkner2009}. Below we directly calculate the mean value of the system energy with its reduced density operator and recognise this as the internal energy of the system:
\begin{equation}\label{energy}
E=\langle H_S \rangle = \frac{\langle p^2 \rangle}{2M} +{\langle V(q) \rangle}
\end{equation}
{We will treat two cases:
\begin{itemize}
\item The Damped Harmonic Oscillator: $V(q)=\frac{M}{2}\omega_0^2{q^2},$
\item The Quantum Free Brownian Particle: $V(q)=0.$
\end{itemize}
}
{In both cases} the mean values are calculated by means of the fluctuation-dissipation theorem:
 \begin{equation}\label{varq}
  \langle q^2 \rangle = \frac{\hbar}{M}f_0(\beta),
  \end{equation}\begin{flushleft}
  
  \end{flushleft}
  
  \begin{equation} \label{varp}
  \langle p^2 \rangle = \hbar Mf_2(\beta),
  \end{equation}
 where
  \begin{equation}\label{FDT}
  f_n(\beta)=\int_{-\infty}^{\infty}\frac{d\omega}{2\pi}\frac{{2\gamma}\, \omega^{n+1}}{(\omega^2 - \omega_0^2)^2 +{4\gamma^{2}}\omega^2}\coth\left(\frac{\hbar\beta\omega}{2}\right).
  \end{equation}
{and $\gamma\equiv\eta/2{M}$}.

  Using {this result} we can numerically calculate the specific heat of the damped harmonic oscillator and the free Brownian particle. {Moreover, equation (\ref{energy}) is very useful if we want to discuss extensivity of our composite system as we see in what follows.}

{Let us study two specific limits of (\ref{energy}), namely the very weakly ($\gamma\ll\omega_{0}$) and the strongly ($\gamma\gg\omega_{0}$) damped cases.} 

{When $\gamma\ll\omega_{0}$ we can easily show that the rational function in the integrand of (\ref{FDT}) becomes a sum of two very well peaked Lorentzians centered at $\pm \omega_{0}$, which implies that the average energy of the particle given by (\ref{energy}) becomes 
\[E=\frac{\hbar \omega_{0}}{2} \coth \frac{\hbar \omega_{0}}{2 k_{B} T}.\]
From this expression we can recover the oscillator ground state energy, $E_{0}=\hbar\omega_{0}/2$, when $T\rightarrow0$ and the equipartition theorem, $E=k_{B}T$, when $k_{B}T \gg \hbar\omega_{0}$. Therefore, we see that in the weakly damped case, which we are considering as representative of the weak coupling regime, the statistical properties of the damped harmonic oscillator are independent of the relaxation frequency $\gamma$ and could be directly obtained from a density operator of the Gibbsian form either in the classical or in the quantum mechanical regime. This means that these statistical properties can be obtained from a reduced density operator which carries no dependence with the coupling to the environment. In other words, the full density operator of the universe is just the product of the density operator of the particle with the density operator of the environment.}

{Although the procedure for $\gamma\gg\omega_{0}$ is not so straightforward, we can still perform the integral in (\ref{FDT}) for $k_{B}T \gg \hbar\omega_{0}$ or $k_{B}T \ll \hbar\omega_{0}$ using partial fraction decomposition for the rational function in its integrand.} 

{In the case of high temperatures the result is the same as for $\gamma\ll\omega_{0}$, namely the equipartition theorem still holds and the density operators of the particle and environment remain separable. However, for $k_{B}T \ll \hbar\omega_{0}$ things quite different.} 

{Let us illustrate this new bahaviour taking $T=0$ as an example. In this case, we can show that
\begin{equation}\label{nonsepenergy}
E_{0}=\frac{\hbar \omega_{0}^{2}}{\pi \gamma} \ln \frac{ 2 \gamma}{\omega_{0}} + \frac{\hbar \gamma}{\pi} \ln \frac{\gamma \Omega}{2\omega_{0}^{2}},
\end{equation}
in full disagreement with the weak coupling regime.} 

{The first term in (\ref{nonsepenergy}) represents the potential energy of the oscillator and becomes negligible in the extreme overdamped limit, $\gamma\ggg\omega_{0}$, because its vanishing prefactor dominates over its logarithmic growth. On the other hand, the second term in (\ref{nonsepenergy}) is the kinetic energy of the oscillator which would diverge logarithmically if not for the presence of the cutoff frequency $\Omega$ in the argument of the logarithmic function. This only happens because in order to write (\ref{FDT}) we rely on the correctness of the Langevin equation (\ref{langevin2}) which is only true for times $t\gg 1/\Omega$. For times shorter than this microscopic time scale the instantaneous form (\ref{langevin2}) is no longer valid because the inertia of the environment comes into play.}

{ Equation (\ref{nonsepenergy}) clearly shows us that its form cannot be obtained from the Gibbs distribution, either classical or quantum mechanical, applied to the oscillator Hamiltonian alone. Actually, as we have already stressed above, this results directly from the reduced density operator  (\ref{redensityop}) which on its turn results from the tracing procedure applied to the Gibbs density operator of the whole universe. In other words, what we are implicitly saying is that the interaction between system and environment cannot be forgotten in the present regime because  it entangles the quantum states of the system and the environment implying in a non-separable equilibrium state. In terms of entropies of the sub-systems it means that they are no longer additive.}  
\begin{figure}
\includegraphics[width=1\linewidth]{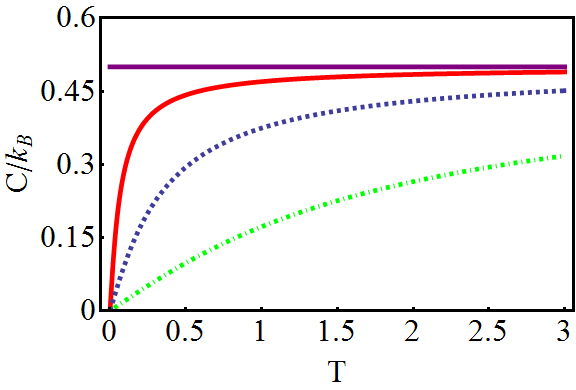}
\caption{ (color online) Numerical calculation of the specific heat of a free Brownian particle for various damping parameters;$\gamma=0$ (purple), $\gamma=0.1$ (red),$\gamma=1$ (dashed blue), and $\gamma=5$ (dotdashed green).}
\includegraphics[width=1\linewidth]{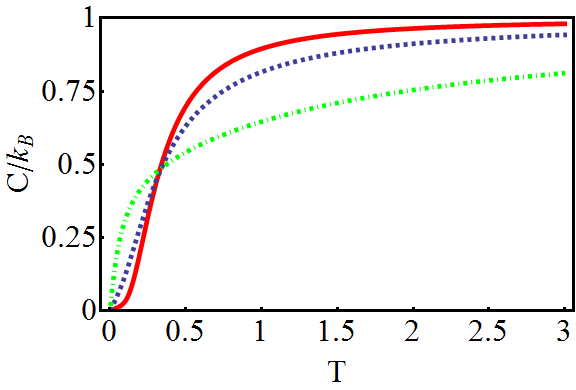}
\caption{ (color online) Numerical calculation of the specific heat of the damped harmonic oscillator for various damping parameters;$\gamma=0.1$ (red),$\gamma=1$ (dashed blue) and $\gamma=5$ (dotdashed green).}
 
\end{figure}
\section{\label{sec:level1b} Non-extensive Entropies Ansatz}
\subsection{\label{sec:level2d}General Remarks}
The above presented scheme requires two ingredients:
\begin{enumerate}
    \item the system-plus-reservoir approach
    \item the choice of the coupling and spectral function
\end{enumerate}
The main question is: Why a bath of noninteracting harmonic oscillators with a coordinate-coordinate coupling and a linear spectral function? In the Brownian particle case, a fair \textit{a posteriori} justification is the fact that the classic equation of motion must be of the Langevin type. A more general justification can be found in\cite{CALDEIRA1983A}.

Another possibility to approach this problem is by means of information theory. {As we have seen above, the coupling to the reservoir (if not weak enough) entangles the system with its environment which implies that the reduced state of the particle, even if it is  Gaussian, is not of a Boltzmann-Gibbs form}. This means that the traditional canonical ensemble maximum entropy scheme cannot give us this density operator. Therefore, in order to capture the features of a dissipative system, we have decided to appeal for the concept of generalised (non-extensive) entropies. 

We address here a model built upon two famous trial entropies: the Rényi\cite{renyi1961} and Tsallis\cite{Tsallis1988} generalised entropies, respectively defined as ; 
\begin{equation}
H_{\alpha}(p)=\frac{1}{1-\alpha}\log[Tr\hat{\rho}^{\alpha}],
\end{equation}
and
\begin{equation}
 S_{q}(p)=\frac{1}{1-q}(Tr\hat{\rho}^{q}-1),
 \end{equation}
where $p$ is the probability distribution associated to the statistical ensemble of the system.
 
These entropies were largely applied to studying diffusion models, and entanglement. Both situations {preserve} some similarities with the strong coupling to a reservoir as in the Brownian motion case.
In a few words, the motivation behind these entropies is to take into account the fact that in the case of interacting systems, the total energy is not the sum of the individual energies, forcing us to move from the single particle picture of the problem to another one in which we can account for all the interactions. From a thermodynamic point of view, this means that entropy is no longer additive and  given that the Boltzmann-Gibbs entropy seems to be always additive, we need a more general form, constrained to recover the additive entropy in some limit.
Applying the maximum entropy formalism using the mean value of energy as a constraint, we surprisingly find a unique form for the density operator for both entropies. This means that:
\begin{equation}
 \langle E \rangle = Tr(\rho_{q/\alpha}H),
 \end{equation}
 together with  the normalisation condition,
 \begin{equation}
 Tr(\rho_{q/\alpha})=1,
 \end{equation}
yield,
\begin{equation}
\rho_q=[1-(1-q)\beta_q H]^{\frac{1}{1-q}}/\mathcal{Z}_q,
\end{equation}
 where $\mathcal{Z}_q$ is the generalised partition function,
 
 \begin{equation}
 \mathcal{Z}_q=Tr(\rho_q).
 \end{equation}
Notice that we have used the parameter $q$ for representing either $q$ itself or $\alpha$ .

The connection with thermodynamics is made in terms of the so-called \textit{q}-log function: 
\begin{equation}
F_q=-k_BT\frac{\mathcal{Z}_q^{1-q} -1}{1-q}=-k_TT\log_q\mathcal{Z}_q,
\end{equation}
where we define the \textit{q}-log as:
\begin{equation}
 \log_q(X) = \frac{X^{1-q}-1}{1-q} 
 \end{equation}
 As a final remark, there is some controversy about the right way to impose the constraint and maximise Tsallis entropy\cite{Tsallis2009}\cite{Tsallis1988}, applying a so-called ``generalised'' mean value using the $q$-moment of the density operator or some kind of normalised distribution with no physical justification. In here, we apply the mean values calculated in the standard way :
 \begin{equation}
\langle A \rangle = Tr(\rho A),
 \end{equation}
 because this form arises directly red from the definition of the density operator and the basic principles of quantum mechanics \cite{merzbacher1998quantum}.
\subsection{\label{sec:level2e}Statement of the Model}
  In the description of dissipative systems, the relaxation constant $\gamma\equiv \eta/2m$ represents all of the dissipative effects of the interaction with the bath, and in the limit $\gamma \rightarrow 0$ we recover the equilibrium correlations of the decoupled system. In nonextensive statistical mechanics, the adjustable parameter $q$ is a measure of how far the system is from extensivity, and in our interpretation the strength of the coupling to the   bath is the cause of nonextensivity. This means that somehow $q$ and $\gamma$ are related, and, in the limit of no damping , $q \rightarrow 1 ,$  we recover the results obtained by the canonical ensemble.
  We now construct a simple model for the reduced density operator: We assume that the energy levels are distributed according to the probability law that emerges from the maximum entropy principle {as} applied to the generalised entropies. So,  our attempt consists in evaluating the density operator for the harmonic oscillator and free particle in these new ensembles. We can justify this model appealing again to the definition of the statistical operator:
  \begin{equation}
       \hat{\rho}=\sum_ip_i\ket{\Psi_i}\bra{\Psi_i},
  \end{equation}
  where $p_i$ is the probability distribution of the statistical ensemble. In our model, this statistical distribution is given by the maximisation of the generalised entropies. In other words, we claim that the physical scenario is reflected upon the ensemble and not on the system itself.
  
  Once we establish a model, the procedure is clear:
  \begin{itemize}
      \item Obtain the generalised density operator
      \item Compute the partition function
      \item Evaluate the thermodynamic quantities
  \end{itemize}
    
    The most reliable method to do this, is the employment of the integral representation\cite{MENDES1999} of the generalised distributions:
    \begin{itemize}
\item $q > 1$
\begin{equation}
\mathcal{Z}_q(\beta)=\frac{1}{\Gamma(\frac{1}{q-1})}\int_{0}^{\infty}dt e^{-t}\mathcal{Z}( -t\beta(1-q))t^{\frac{1}{q-1} -1},
\end{equation}
\item $q < 1$
\begin{equation}
\mathcal{Z}_{q}(\beta)=\frac{i}{2\pi}\Gamma(\frac{2-q}{1-q})\oint_{C}dt e^{-t}\mathcal{Z}(\beta(1-q))t^{-\frac{2-q}{1-q}},
\end{equation}
\end{itemize}
where, $\mathcal{Z}$ is the canonical partition function, with a rescaled temperature; $\beta \rightarrow \beta (q-1).$

These representations apply for both Rényi and Tsallis distributions and act like integral transforms connecting the canonical ensemble to the Rényi-Tsallis ensemble. From the partition function of the harmonic oscillator and free particle we obtain their parameter dependent versions as
\begin{equation}
\mathcal{Z}_q(\beta)=\frac{{1}}{{\beta(q-1)}}^{\frac{1}{q-1}}\zeta\left[\frac{1}{q-1},\frac{1}{2}+\frac{{1}}{{\beta(q-1)}}\right],
\end{equation}
{for the harmonic oscillator, and, for the free particle,}
\begin{eqnarray}
 Z_q(\beta) &=& \int_{-L/2}^{L/2}\rho_{q}(x,x,\beta)dx \\ &=& L\left [\frac{m}{2\pi(q-1)\beta\hbar^2} \right]^{\frac{1}{2}}\frac{\Gamma\left(\frac{1}{q-1}-\frac{1}{2}\right )}{\Gamma\left(\frac{1}{q-1}\right )},
 \end{eqnarray}
when $q>1$, and
\begin{equation}
 Z_q(\beta) = L\left [\frac{m}{2\pi(1-q)\beta\hbar^2} \right]^{\frac{1}{2}}\frac{\Gamma\left(\frac{2-q}{1-q}\right )}{\Gamma\left(\frac{2-q}{1-q}+\frac{1}{2}\right )},
 \end{equation}
for $q< 1$. In both cases, $L$ is the size of the normalisation box.
 
 In the case of the harmonic oscillator, the result is valid only in the region $1<q<3$. At first sight this limitation has no physical meaning and seems to come from the properties of the Gamma function. Care must be taken in handling the results for $q<1$ in the free particle case because this parameter cannot reach negative values for two reasons:
 \begin{enumerate}
     \item Rényi entropy is not defined for negative values of the parameter {$q$}.
     \item Tsallis entropy is defined for negative values of $q$, however it changes the concavity of the entropy and the thermodynamic meaning of some quantities.
 \end{enumerate}
 We now compute the specific heat in this model in order to compare it with the harmonic oscillator bath results.  
\subsection{\label{sec:level2f} Thermodynamic Properties}

From the above section, we can conclude that the range of applicability of this ansatz is limited, and corresponds to small values of $\gamma$, or weak dissipation. However,  we need to further investigate the results and also test their meaning  within the prescription of thermodynamics.  As in the harmonic bath case we use the specific heat as a thermodynamic parameter. Therefore,
\begin{equation}
F_q=-k_{{B}}T\frac{\mathcal{Z}_q^{1-q} -1}{1-q}=-k_{{B}}T \log_q\mathcal{Z}_q,
\end{equation}
{and}
\begin{equation}
C_q=-T\frac{d^2F_q}{d^2T}.
\end{equation}

In the harmonic oscillator case, we performed a numerical calculation and found a specific heat with an anomalous behaviour as shown in \autoref{fig:5}, that is clearly incompatible with the second law of thermodynamics, and so  we cannot make any such a   comparison in the region of existence of $\mathcal{Z}_q$.
\begin{figure}
\centering
    \includegraphics[width=1\linewidth]{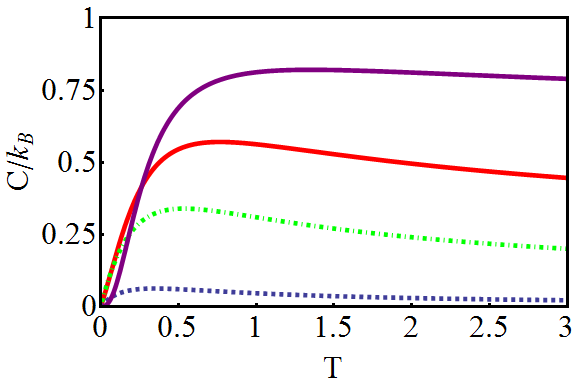}
\caption{ \label{fig:5}(color online) Specific heat of the  harmonic oscillator in the Rényi-Tsallis Ensemble for  $q=1.1$ (purple), $q=1.3 $(red), $q=1.5$(dotdashed green) and $q=1.8$ (dashed blue). In this case there is convergence only in the range $1< q < 3$, and this has no physical interpretation. However, in all cases we see a disagreement with the second law of thermodynamics, and , therefore, these are also unphysical solutions.} 
\end{figure}
To calculate the specific heat of the free particle is an easy task, and we have,
\begin{itemize}
\item $q>1$
\begin{equation}
C_q=2^{\frac{1}{2}(q-5)}\pi^{\frac{1}{2}(q-1)}(q-3)\left(\sqrt[]{\frac{T}{q-1}}\frac{\Gamma(\frac{1}{q-1}-\frac{1}{2})}{\Gamma(\frac{1}{q-1})}\right)^{1-q}
\end{equation}
\item $q<1$
\begin{equation}
C_q=2^{\frac{1}{2}(q-5)}\pi^{\frac{1}{2}(1-q)}(q-3)\left(\sqrt[]{\frac{T}{1-q}}\frac{\Gamma(\frac{1}{1-q}+1)}{\Gamma(\frac{1}{1-q}+\frac{3}{2})}\right)^{1-q}
\end{equation}
\end{itemize}
\begin{figure}[h!]
    \centering
    \includegraphics[width=1\linewidth]{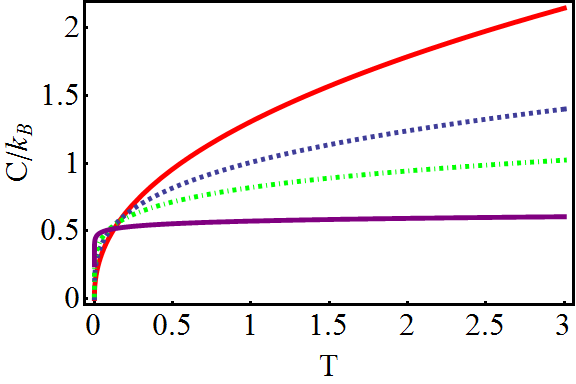}
    \caption{\label{fig:3}(color online) Specific heat of redthe free particle in red the Rényi-Tsallis Ensemble for $(q<1);$ $q=0.1$ (red), $q=0.4$ (dotdashed green), $q=0.6$ (dashed blue), and $q=0.9$ (purple). These obey all the laws of thermodynamics but show little similarity with the free Brownian motion.} 
\includegraphics[width=1\linewidth]{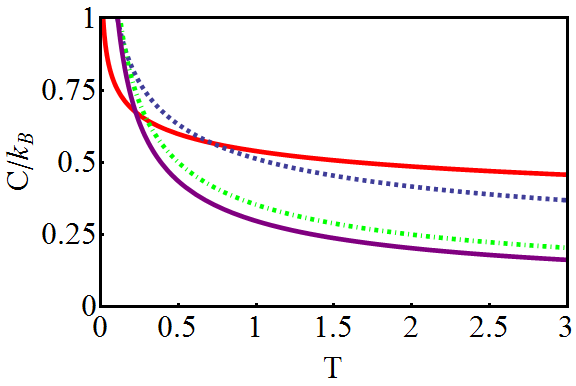}
\caption{\label{fig:4}(color online) Specific heat of the free particle in the Rényi-Tsallis ensemble for $(q>1)$, $q=1.3$ (red), $q=1.6$ (dotdashed green), $q=2$ (dashed blue) and $q=2.1$ (purple). These results do not obey the second and third laws of thermodynamics and must be disregarded as unphysical solutions.}
\end{figure}
From \autoref{fig:3}, we see that only the case $q<1$ has physical meaning, because a decaying heat capacity is forbidden\cite{Hilbert2014} , and, on top of that, the result for $q>1$ does not obey the third law of thermodynamics\cite{Ingold2012}. For  $0<q<1$ we have  a very limited region of validity, and using $q$ as a fit parameter we found a numerical correlation between the adjustable parameter and the damping parameter. 
\begin{figure}[h!]
\includegraphics[width=1 \linewidth]{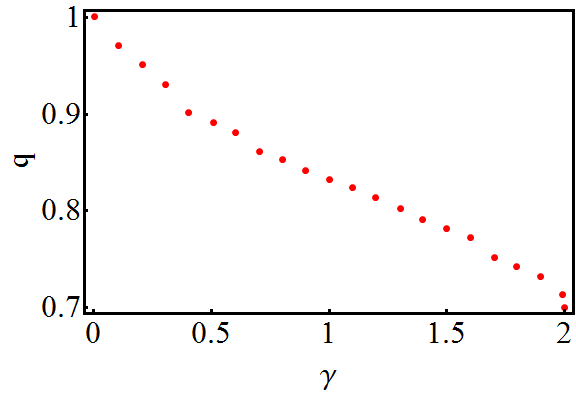}
\caption{\label{fig:6} Numerical comparison between the adjustable parameter and the damping parameter}

\end{figure}

If $\gamma=0$ we clearly recover the $q=1$ case, as the first test of the model. However, this relation has an upper bound in the free region and a change of concavity at the transition from  the weak damping ($\gamma < 1$) to the strong damping ($\gamma > 1$) regimes. We expected to find different damping regimes for different $q$ values, and furnish some physical meaning to $q$. Our results show that the thermodynamic functions obtained in this way , for these particular examples, act as a very limited fitting ansatz and are not a generalised solution as claimed in the literature. Moreover, numerical calculations do not show any quantitative agreement between the two methods  as shown in \autoref{fig:7}

\begin{figure}[h!]

\includegraphics[width=1\linewidth]{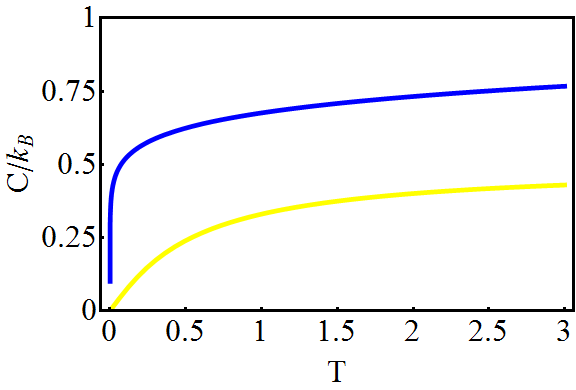}
\caption{\label{fig:7} Typical behaviour of the specific heat calculated with the harmonic oscillator heat bath, $\gamma=1.5$ (blue), and the non-extensive entropy prescription,$q=0.77$ (purple).}

\end{figure}

In every scenario, we found a specific heat that decays with temperature, which is at variance with   the second law and must also be discarded.
\section{\label{sec:level1c}Conclusions}
In this work, after reviewing the main aspects of the system-plus-reservoir method to dissipative systems, we indicate how the  thermodynamic properties of simple dissipative systems should be computed. Then, relying on the possibility of describing the non-extensivity of the entropy of the composite universe, we constructed a model based on the maximisation of the generalised entropies of Rényi and Tsallis, and by using the connections with thermodynamics we tried to insert quantum Brownian motion within this novel picture. 

{ Our results show the existence of some pitfalls in the thermodynamic properties reached in this way, owing to the presence of regimes in which the thermodynamic laws are no longer respected as the decaying specific heat shown in \autoref{fig:5} and \autoref{fig:4}, that disagrees with the second law of thermodynamics. It is important to make clear that the claim that entropy always grows with temperature is based only on the second law and not on any particular form of ensemble one chooses.  Moreover, even in those limits where the general behaviour of the two methods are physically sound, the quantitative agreement is very poor (as one can see from  \autoref{fig:6} and \autoref{fig:7}) being unable to reproduce the thermodynamic behaviour of the Brownian particle.}

We conclude that the system-plus-reservoir approach is more general and complete to describe the thermodynamics of dissipative systems and the employment of generalised entropies in the thermodynamic context should be avoided. The main point is the fact that those entropies can be used as fine tools to measure information does not imply that they will be equally reliable in the thermodynamic realm. In  this sense we argue that these entropies have limited applicability to thermodynamics in the strong coupling regime, showing that it is not a proper generalisation of thermostatistics.
\section{\label{sec:level1d}Acknowledgements} 
 A. O. C. wishes to acknowledge financial support from ``Conselho Nacional de Desenvolvimento Científico e Tecnológico (CNPq)'' and ``Fundação de Amparo à Pesquisa no Estado de São Paulo (FAPESP)'' through the initiative ``Instituto Nacional de Ciência e Tecnologia em Informação Quântica (INCT-IQ)'' whereas L.M.M.D  has been supported by  ``Coordenação de Aperfeiçoamento de Pessoal de Nível Superior (CAPES)''. Both of us are deeply indebted to M. Campisi and T.V. Acconcia for enriching discussions and references.

 \bibliography{refend}
 
\end{document}